\documentclass[         %
aps,                    
jpa,                    
showpacs,               
nofootinbib,            
showkeys,               %
preprintnumbers,        %
floatfix]               
{revtex4}               
\tighten
\usepackage{graphicx}

\begin{document}

\title{Symmetries of the Neutrino Interactions}

\pacs{02.20.-a, 11.30.Fs, 14.60.Pq, 26.30.Jk}
\keywords      {Neutrino mass and mixings, symmetries in neutrino physics, core-collapse supernova, 
collective neutrino oscillations}

\author{A.B. Balantekin}
\address{Physics Department, University of Wisconsin, Madison WI 53706 USA}

\begin{abstract}
Symmetries in neutrino physics are explored using  analogies to fermion pairing in many-body systems. 
In particular, the SO(5) symmetry of the most general neutrino mass Hamiltonian with both Dirac and 
Majorana mass terms as well as the invariants of certain limits of the collective neutrino Hamiltonian are 
discussed. The latter Hamiltonian finds applications in the Early Universe and core-collapse supernovae. 
\end{abstract}

\maketitle


\section{Introduction}

Recent years have seen an increasingly intense activity in neutrino physics. Not only has neutrino mixing been  
experimentally established, but also the three mixing angles have been measured with reasonable accuracy.   
Direct neutrino mass measurements have not yet reached down to the neutrino mass scale chosen by the Nature, but differences 
between squares of masses are deduced from oscillation experiments again with reasonable accuracy 
despite the fact that 
the mass hierarchy remains an open question. 

Symmetries play a crucial role is constraining the behavior of physical systems. In this contribution I describe 
how symmetry concepts apply to neutrino physics using analogies to fermion pairing in many-body systems. 
One use of this analogy, namely the SO(5) symmetry which connects the Dirac and Majorana masses of neutrinos with the see-saw mechanism 
is discussed in the next section. 

Since our current knowledge of 
neutrino parameters is rather robust, we can now address applications of neutrino physics in astrophysical 
settings with more confidence. One such application is the self-interacting neutrino gas present in the 
core-collapse supernovae and the Early Universe. Symmetries, the integrability and associated constants 
of motion of the Hamiltonian 
describing such a many-neutrino gas is described in the section after the next one. 

\section{SO(5) algebra and the neutrino mass}

Both the Dirac mass term
\begin{equation}
H_m^D= m_D\int d^3{x} (\bar{\psi}_L\psi_R+h.c.)
\end{equation}
as well as the left- and right-handed Majorana mass terms 
\begin{equation}
H_m^M = H_m^L+H_m^R = \frac{1}{2}m_L\int d^3{x}(\bar{\psi}_L\psi_L^c+h.c)
+\frac{1}{2}m_R\int d^3{x} (\bar{\psi}_R\psi_R^c+h.c), 
\end{equation}
where the charge-conjugate spinor is defined as $
\psi^C= C \bar{\psi}^T$, 
can be written in terms of the generators of an SO(5) algebra \cite{Balantekin:2000qt}. 
The generators of the $SU(2)_D$ subalgebra 
of this SO(5) are 
\begin{equation}
D_- = \int d^3{x}(\bar{\psi}_R\psi_L) = D_+^\dagger, \> 
D_0=\frac{1}{2}\int d^3{x}(\psi_L^{\dag}\psi_L-\psi_R^{\dag}\psi_R) ,
\end{equation}
whereas the generators of the $SU(2)_L \times SU(2)_R$ subalgebra are 
\begin{equation}
L_+ = \frac{1}{2}\int d^3{x}(\bar{\psi}_L\psi_L^c) = L_-^\dagger, 
L_0=\frac{1}{4}\int d^3{x}(\psi_L^{\dag}\psi_L-\psi_L\psi_L^{\dag})
\end{equation}
and 
\begin{equation}
R_+=\frac{1}{2}\int d^3{x} (\bar{\psi_R^c}\psi_R) = R_-^\dagger,  
R_0=\frac{1}{4}\int d^3{x}(\psi_R \psi_R^{\dag}-\psi_R^{\dag}\psi_R) .
\end{equation}
Note that $D_0$ is not an independent operator, but the sum of $L_0$ and $ R_0$. 
With the addition of two more operators, 
\begin{equation}
\label{6}
A_+ = \int d^3{x} \left[ -\psi_L^T C\gamma_0\psi_R \right], \> 
 A_-= \int
d^3{x} \left[ \psi_R^{\dag} \gamma_0 C (\psi_L^{\dag})^T\right], 
\end{equation}
one has the ten generators of the $SO(5)$ pairing algebra. 

To explore the physical meaning of this SO(5) algebra, one can consider 
the Pauli-G\"{u}rsey transformation \cite{PG} 
\begin{equation}
\psi\rightarrow\psi^{\prime}=a\psi+b\gamma_5\psi^c,  \>\> |a|^2+|b|^2=1 ,
\end{equation} 
which describes particle-antiparticle mixing. The two operators $A_+$ and $A_-$ of Eq. (\ref{6}) along with 
$A_0 \equiv R_0 - L_0$ generate another SU(2) algebra that we call $SU(2)_{\rm PG}$.  The 
most general element of the associated $SU(2)_{\rm PG}$ group is 
\begin{equation}
\hat{U}=e^{-\tau^* A_-}e^{-log(1+|\tau|^2)A_0}e^{\tau A_+}e^{i\varphi
A_0}
\end{equation} 
under which the field $\psi$ transforms as 
\begin{equation}
\psi \rightarrow \psi' =
\hat{U}\psi\hat{U}^{\dag}=\frac{e^{i\varphi/2}}{\sqrt{1+|\tau|^2}}
[\psi-\tau^* \gamma_5 \psi^c] .
\end{equation}
Clearly this is a Pauli-G\"{u}rsey transformation with
\[
a=\frac{e^{i\varphi/2}}{\sqrt{1+|\tau|^2}} \;\; ,\;\; b=\frac{-\tau^*
e^{i\varphi/2}}{\sqrt{1+|\tau|^2}} .
\]
The most general neutrino mass Hamiltonian: 
\[
H_{m}=m_D(D_++D_-)+m_L(L_++L_-)+m_R(R_++R_-) 
\]
sits in the $SO(5)/SU(2)_{\rm PG} \times U(1)_{\chi}$ coset where
$U(1)_{\chi}$ is generated by $D_0=L_0+R_0$.  Under the Pauli-G\"ursey transformation 
Dirac mass terms would transform into a mixture of Dirac and Majorana mass terms, providing 
an algebraic framework for the see-saw mechanism. 

Although the discussion here pertains only to one single flavor, it is also possible to generalize these algebraic 
arguments to multiple neutrino flavors \cite{Ozturk:2001zn}. 

\section{Symmetries and integrability of the self-interacting neutrino gas}

We first consider only two flavors of neutrinos: electron neutrino, $\nu_e$, and 
another flavor, $\nu_x$. Introducing the creation and annihilation operators for a neutrino 
with three momentum ${\bf p}$, we can write down the generators of the neutrino flavor isospin algebras   
\cite{Balantekin:2006tg}: 
\begin{eqnarray}
J_+({\bf p}) &=& a_x^\dagger({\bf p}) a_e({\bf p}), \> \> \>
J_-({\bf p})=a_e^\dagger({\bf p}) a_x({\bf p}), \nonumber \\
J_0({\bf p}) &=& \frac{1}{2}\left(a_x^\dagger({\bf p})a_x({\bf p})-a_e^\dagger({\bf p})a_e({\bf p})
\right). \label{su2}
\end{eqnarray}
The integrals of these operators over all possible values of momenta generate the global flavor 
isospin algebra. Using the operators in Eq. (\ref{su2}) 
the Hamiltonian for a neutrino propagating through matter takes the form  
\begin{equation}
\label{msw}
 H_{\nu} = \int d^3{\bf p} \frac{\delta m^2}{2p} \left[
\cos{2\theta} J_0({\bf p}) + \frac{1}{2} \sin{2\theta}
\left(J_+({\bf p})+J_-({\bf p})\right) \right] -  \sqrt{2} G_F \int d^3{\bf p} 
\> N_e \>  J_0({\bf p}).  
\end{equation}
In Eq. (\ref{msw}), the first integral represents the neutrino mixing and the second integral 
represents the neutrino forward scattering off the background matter. In writing this equation 
a term proportional to identity is omitted as such terms do not contribute to the neutrino oscillations. 
Neutrino-neutrino interactions  (see references \cite{Pantaleone:1992eq} through \cite{Saviano:2012yh} 
are described by the Hamiltonian 
\begin{equation}
\label{nunu}
H_{\nu \nu} = \sqrt{2} \frac{G_F}{V} \int d^3{\bf p} \> d^3{\bf q} \>  (1-\cos\vartheta_{\bf pq}) \> {\bf
J}({\bf p}) \cdot {\bf J}({\bf q}) ,
\end{equation}
where $\vartheta_{\bf pq}$ is the angle between neutrino momenta {\bf p} and {\bf q} and V 
is the normalization volume.  Note that the presence of the $(1-\cos\vartheta_{\bf pq}) $ term in 
the integral above is crucial to recover the effects of the Standard Model weak  
interaction physics in the most general situation\footnote{For a recent discussion of the impact of the physics 
beyond the Standard Model 
on matter-enhanced neutrino oscillations see \cite{Balantekin:2011ft} and references therein.}. 
Note that in the extremely idealized case of 
isotropic neutrino distribution and a very large number of neutrinos, this term may average to a 
constant and the neutrino-neutrino interaction 
Hamiltonian simply reduces to the Casimir operator of the global SU(2) algebra. This is unlikely 
for core-collapse supernovae, but is the case for the Early Universe 
\cite{Kostelecky:1993ys,Abazajian:2002qx}. 

The discussion above pertains to a gas comprised of neutrinos only. Inclusion of antineutrinos 
as well requires introduction of a second set of SU(2) algebras. Similarly for three flavors two 
sets of SU(3) algebras 
are needed \cite{Sawyer:2005jk}. Both extensions are straightforward, but tedious. 

Defining the auxiliary vector quantity 
\begin{equation}
\hat{B} = (\sin2\theta,0,-\cos2\theta), 
\end{equation}
the total Hamiltonian with two flavors,containing one- and two-body interaction terms, can be written as
\begin{equation}
\label{total}
\hat{H}_{\mbox{\tiny total}} = H_{\nu} + H_{\nu \nu} 
= \left(
\sum_p\frac{\delta m^2}{2p}\hat{B}\cdot\vec{J}_p  - \sqrt{2} G_F 
N_e  J_p^0  \right) 
+ \frac{\sqrt{2}G_{F}}{V}\sum_{\mathbf{p},\mathbf{q}}\left(1- 
\cos\vartheta_{\mathbf{p}\mathbf{q}}\right)\vec{J}_{\mathbf{p}}\cdot\vec{J}_{\mathbf{q}}  
\end{equation} 

The evolution operator for the system represented by the Hamiltonian in Eq. (\ref{total})  
\begin{equation}
i\frac{\partial U}{\partial t} = \left( H_{\nu} + H_{\nu \nu} \right)  U ,
\end{equation}
can be approximately evaluated using the stationary phase approximation to its path integral 
representation \cite{Balantekin:2006tg}. This is equivalent to reducing the Hamiltonian 
$H_{\nu \nu}$,  to a 
one-body one in an RPA-like approximation. In this approximation the product of two commuting 
operators  $\hat{\cal O}_1$ and $\hat{\cal O}_2$ is approximated as 
\begin{equation}
\label{16a}
\hat{\cal O}_1 \hat{\cal O}_2 \sim 
\hat{\cal O}_1 \langle \hat{\cal O}_2 \rangle + \langle \hat{\cal O}_1 \rangle \hat{\cal O}_2 -
\langle \hat{\cal O}_1 \rangle \langle \hat{\cal O}_2 \rangle ,
\end{equation}
where the expectation values should be calculated with respect to a well-chosen state $|\Psi\rangle$ which satisfies 
the condition $ \langle \hat{\cal O}_1  \hat{\cal O}_2 \rangle = \langle \hat{\cal O}_1 \rangle \langle 
\hat{\cal O}_2\rangle~$. One then obtains the single-angle Hamiltonian 
\begin{equation}
\label{17}
\hat{H}\sim\hat{H}^{\mbox{\tiny RPA}} =  \sum_p\omega_{p}\hat{B}\cdot\vec{J}_p
+\vec{P}\cdot\vec{J}. 
\end{equation}
In writing Eq. (\ref{17})  matter terms are neglected and the polarization vector $\vec{P}$ was defined as
\begin{equation}
\label{18}
\vec{P}_{\mathbf{p},s}=2\langle\vec{J}_{\mathbf{p},s}\rangle . 
\end{equation}
Using the SU(2) coherent states associated with the flavor isospin in calculating the operator averages 
in the above equations yields the standard reduced collective neutrino Hamiltonian, widely used in the 
literature \cite{Balantekin:2006tg}. 

Both the full Hamiltonian and its one-body reduction possess an $SU(N)_f$ rotation 
symmetry in the neutrino flavor space \cite{Balantekin:2006tg,Duan:2008fd,Balantekin:2009dy}. 
Such a complex nonlinear system could presumably exhibit further symmetries. A few of the conserved 
quantities in collective neutrino oscillations were already noted in the literature 
\cite{Raffelt:2007cb,Duan:2007mv}. To look for further invariants, defining $\mu=\frac{\sqrt{2}G_{F}}{V}$, 
$\tau=\mu t$, and  $ \omega_{p}=\frac{1}{\mu}\frac{\delta m^{2}}{2p}$ \cite{Pehlivan:2010zz},  one can write the 
Hamiltonian of Eq. (\ref{total}) in the single-angle approximation as
\begin{equation}
\label{18}
\hat{H} = \sum_p\omega_{p}\hat{B}\cdot\vec{J}_p
+\vec{J}\cdot\vec{J} .
\end{equation}
In writing Eq. (\ref{18}), it was assumed that neutrino-neutrino interaction term is dominant and hence the matter 
terms are ignored. This Hamiltonian preserves the \emph{length of each spin}
\begin{equation}
\label{20}
\hat{L}_p=\vec{J}_{p}\cdot\vec{J}_{p}~,
\qquad\qquad \left[ \hat{H}, \hat{L}_p\right]=0~,
\end{equation}
as well as the \emph{total spin component} in the direction of the "external magnetic field", $\hat{B}$  
\begin{equation}
\hat{C}_0 =\hat{B}\cdot\vec{J}~, \qquad\qquad\quad \left[\hat{H},\hat{C}_0\right]=0~ . 
\end{equation}
The conservation law depicted in Eq. (\ref{20}) is nothing but the conservation of the total number of neutrinos 
with a given momentum as neither neutrino mixing nor coherent forward scattering of neutrinos off one another 
change that number. 

The Hamiltonian in Eq. (\ref{18}) is similar to the reduced BCS Hamiltonian of the many-body theory. 
(However, one should note that the sign of the pairing term is opposite to that in the BCS Hamiltonian). 
One can exploit this duality to uncover its symmetries. Eigenstates of the reduced BCS Hamiltonian were 
written by Richardson using a Bethe ansatz in \cite{Richardson1} and later generalized by 
Gaudin \cite{Gaudin1,Gaudin2}. Since the BCS Hamiltonian considered by Richardson is integrable, there are 
constants of motion associated with it \cite{yuzb}. Using this analogy one can write down the constants of motion of the collective neutrino Hamiltonian in Eq. (\ref{total}) as \cite{Pehlivan:2011hp}
\begin{equation}
\hat{h}_{p} = \hat{B}\cdot\vec{J}_p+2\sum_{q\left(\neq p\right)}\frac{\vec{J}_{p}\cdot\vec{J}_{q}}{\omega_{p}-\omega_{q}}.
\end{equation}
The individual neutrino spin-lengths discussed above, $\hat{L}_p$,  are independent invariants. They are set by the initial 
conditions and are not changed by the evolution of the system under the collective Hamiltonian. 
However one has 
$
\hat{C}_0=\sum_{p}\hat{h}_{p}$. 
The Hamiltonian itself is also a linear combination of these invariants: 
\[
\hat{H}=\sum_{p}w_{p}\hat{h}_{p}+\sum_{p} \hat{L}_{p}~.
\]

Including antineutrinos, the conserved quantities for each neutrino energy mode $p$ take the form 
\begin{equation}
\hat{h}_{p} = \hat{B}\cdot\vec{J}_p+2\sum_{q\left(\neq p\right)}\frac{\vec{J}_{p}\cdot\vec{J}_{q}}{\omega_{p}-\omega_{q}}+2\sum_{\bar{q}}\frac{\vec{J}_{p}\cdot\vec{\tilde{J}}_{\bar{q}}}{\omega_{p}-\omega_{\bar{q}}}, 
\end{equation}
where for antineutrinos we defined $ \omega_{\bar{p}}=-\frac{1}{\mu}\frac{\delta m^2}{2\bar{p}}$. 
Conserved quantities $\hat{h}_{\bar{p}}$ for different antineutrino energy modes are 
\begin{equation}
\hat{h}_{\bar{p}} =\hat{B}\cdot\vec{\tilde{J}}_p+2\sum_{\bar{q}\left(\neq\bar{p}\right)}\frac{\vec{\tilde{J}}_{\bar{p}}\cdot\vec{\tilde{J}}_{\bar{q}}}{\omega_{\bar{p}}-\omega_{\bar{q}}}+2\sum_{q}\frac{\vec{\tilde{J}}_{\bar{p}}\cdot\vec{J}_{q}}{\omega_{\bar{p}}-\omega_{q}}~.
\end{equation}
The invariants of the one-body Hamiltonian of Eq. (\ref{17}) can be written from those as 
\begin{equation}
I_p=2\langle\hat{h}_{p}\rangle =\hat{B}\cdot\vec{P}_p+\sum_{q\left(\neq p\right)}\frac{\vec{P}_{p}\cdot\vec{P}_{q}}{\omega_{p}-\omega_{q}}+\sum_{\bar{q}}\frac{\vec{P}_{p}\cdot\vec{\tilde{P}}_{\bar{q}}}{\omega_{p}-\omega_{\bar{q}}}
\end{equation}
and 
\begin{equation}
I_{\bar{p}}=2\langle\hat{h}_{\bar{p}}\rangle = \hat{B}\cdot\vec{\tilde{P}}_{\bar{p}}+\sum_{\bar{q}\left(\neq\bar{p}\right)}\frac{\vec{\tilde{P}}_{\bar{p}}\cdot\vec{\tilde{P}}_{\bar{q}}}{\omega_{\bar{p}}-\omega_{\bar{q}}}+\sum_{q}\frac{\vec{\tilde{P}}_{\bar{p}}\cdot\vec{P}_{q}}{\omega_{\bar{p}}-\omega_{q}} .
\end{equation}
It was shown that existence of such invariants could lead to collective neutrino oscillations \cite{Raffelt:2011yb}. 

As in the BCS theory, the one-body Hamiltonian of Eq. (\ref{17}) does not conserve particle number. Particle 
number conservation can be enforced by introducing a Lagrange multiplier: 
\begin{equation}
\hat{H}^{\mbox{\tiny RPA}} \rightarrow \hat{H}^{\mbox{\tiny RPA}}+\omega_c\hat{J}^0.  
\end{equation}
Diagonalization of this Hamiltonian gives rise to the phenomena called spectral split or swapping in the 
neutrino energy spectra \cite{Raffelt:2007cb,Duan:2007bt,Dasgupta:2009mg,Galais:2011gh} with 
the Lagrange multiplier playing the role of the swap frequency. 
  
The fact that invariants of the full Hamiltonian are also invariants of the one-body Hamiltonian of Eq. (\ref{17}) 
when they are properly linearized provides confidence in the aptness of the linearization procedure itself. 
One should also note that 
another linearization procedure has been used to carry out flavor-stability analysis of dense neutrino streams 
\cite{Banerjee:2011fj}. 
  
\section{Conclusions}

Neutrinos play an important role in astrophysical settings, including core-collapse supernovae  
\cite{Balantekin:2003ip}.  Core-collapse supernovae are one of the possible sites for the r-process 
nucleosynthesis. Yields of r-process nucleosynthesis are determined by the electron fraction, or equivalently 
by the neutron-to-proton ratio, n/p. Interactions of the neutrinos and antineutrinos streaming out of the 
core both with nucleons and seed nuclei determine the n/p ratio. Hence it is crucial to understand neutrino 
properties, neutrino interactions, and symmetries of those interactions in a many-body environment.  

We examined the many-neutrino gas  both from the exact many-body perspective and from the point of view of an effective one-body description formulated with the application of the RPA method. In the limit of the single angle approximation, 
we showed that both the many-body and the RPA pictures possess many constants of motion manifesting the existence of associated dynamical symmetries in the system. 
The existence of such 
constants of motion offer practical ways of extracting information even from exceedingly complex systems. Even when the symmetries which guarantee their existence is broken, they usually provide a convenient set of variables which behave in a relatively simple manner depending on how drastic the symmetry breaking factor is.
The existence of such invariants naturally lead to associated  collective modes in neutrino oscillations. It should be emphasized that existence of invariants does not obviate numerical analysis as the stability of such collective behavior still needs to be numerically studied.

Whether there are invariants associated with the more realistic multi-angle collective neutrino Hamiltonian is an open question.  Note that even the multi-angle picture may need to be modified because of the presence of 
neutrinos that undergo direction-changing scattering outside of the neutrinosphere \cite{Cherry:2012zw}. 
  

\section*{Acknowledgments}
This work was supported in part
by the U.S. National Science Foundation Grants No. 
PHY-0855082 and PHY-1205024.
and
in part by the University of Wisconsin Research Committee with funds
granted by the Wisconsin Alumni Research Foundation. 
I also thank the Center for Theoretical Underground Physics and Related Areas (CETUP* 2012) in South Dakota for its hospitality and for partial support during the completion of this work






\begin{thebibliography}{99}

\bibitem{Balantekin:2000qt} 
  A.~B.~Balantekin and N.~Ozturk,
  \emph{Phys.\ Rev.\ D} \textbf{62}, 053002 (2000)
  [hep-th/0003260].

\bibitem{PG}
W. Pauli, Nuovo Cimento {\bf 6}, 204 (1957); F. G\"ursey, Nuovo Cimento 
{\bf 7}, 411 (1957). 

\bibitem{Ozturk:2001zn} 
  N.~Ozturk,
  Phys.\ Scripta T {\bf 93}, 41 (2001)
  [hep-th/0102196].

\bibitem{Balantekin:2006tg} 
  A.~B.~Balantekin and Y.~Pehlivan,
  \emph{J.\ Phys.\ G} \textbf{34}, 47-66 (2007)
  [astro-ph/0607527].

\bibitem{Pantaleone:1992eq}
  J.~T.~Pantaleone,
  Phys.\ Lett.\ B {\bf 287}, 128 (1992); 
  Phys.\ Rev.\ D {\bf 46}, 510 (1992).

\bibitem{Qian:1994wh}
  Y.~Z.~Qian and G.~M.~Fuller,
  Phys.\ Rev.\ D {\bf 51}, 1479 (1995)
  [arXiv:astro-ph/9406073].

\bibitem{Pastor:2002we}
  S.~Pastor and G.~Raffelt,
  Phys.\ Rev.\ Lett.\  {\bf 89}, 191101 (2002)
  [arXiv:astro-ph/0207281].

\bibitem{Friedland:2003dv}
  A.~Friedland and C.~Lunardini,
  Phys.\ Rev.\ D {\bf 68}, 013007 (2003)
  [arXiv:hep-ph/0304055].

\bibitem{Balantekin:2004ug}
  A.~B.~Balantekin and H.~Yuksel,
  New J.\ Phys.\  {\bf 7}, 51 (2005)
  [arXiv:astro-ph/0411159].

\bibitem{Duan:2005cp}
  H.~Duan, G.~M.~Fuller and Y.~Z.~Qian,
  Phys.\ Rev.\  D {\bf 74}, 123004 (2006)
  [arXiv:astro-ph/0511275]

\bibitem{Hannestad:2006nj}
  S.~Hannestad, G.~G.~Raffelt, G.~Sigl and Y.~Y.~Y.~Wong,
  Phys.\ Rev.\  D {\bf 74}, 105010 (2006)
  [Erratum-ibid.\  D {\bf 76}, 029901 (2007)]
  [arXiv:astro-ph/0608695].
  
\bibitem{Raffelt:2007cb}
  G.~G.~Raffelt and A.~Y.~Smirnov,
  Phys.\ Rev.\  D {\bf 76}, 081301 (2007)
  [Erratum-ibid.\  D {\bf 77}, 029903 (2008)]
  [arXiv:0705.1830 [hep-ph]]; 
  Phys.\ Rev.\  D {\bf 76}, 125008 (2007) 
  [arXiv:0709.4641 [hep-ph]]. 

\bibitem{Friedland:2006ke}
  A.~Friedland, B.~H.~J.~McKellar and I.~Okuniewicz,
  %
  Phys.\ Rev.\ D {\bf 73}, 093002 (2006)
  [arXiv:hep-ph/0602016].
  
\bibitem{Duan:2007mv} 
  H.~Duan, G.~M.~Fuller, J.~Carlson and Y.~-Z.~Qian,
  Phys.\ Rev.\ D {\bf 75}, 125005 (2007)
  [astro-ph/0703776].
  
\bibitem{Fogli:2007bk} 
  G.~L.~Fogli, E.~Lisi, A.~Marrone and A.~Mirizzi,
  JCAP {\bf 0712}, 010 (2007)
  [arXiv:0707.1998 [hep-ph]].

\bibitem{Chakraborty:2009ej} 
  S.~Chakraborty, S.~Choubey, S.~Goswami and K.~Kar,
  JCAP {\bf 1006}, 007 (2010)
  [arXiv:0911.1218 [hep-ph]].
  
\bibitem{Kneller:2010sc} 
  J.~P.~Kneller and C.~Volpe,
  Phys.\ Rev.\ D {\bf 82}, 123004 (2010)
  [arXiv:1006.0913 [hep-ph]].

\bibitem{Dasgupta:2011jf} 
  B.~Dasgupta, E.~P.~O'Connor and C.~D.~Ott,
  Phys.\ Rev.\ D {\bf 85}, 065008 (2012)
  [arXiv:1106.1167 [astro-ph.SR]].

\bibitem{Mirizzi:2011tu} 
  A.~Mirizzi and P.~D.~Serpico,
  Phys.\ Rev.\ Lett.\  {\bf 108}, 231102 (2012)
  [arXiv:1110.0022 [hep-ph]].
  
\bibitem{Saviano:2012yh} 
  N.~Saviano, S.~Chakraborty, T.~Fischer and A.~Mirizzi,
  Phys.\ Rev.\ D {\bf 85}, 113002 (2012)
  [arXiv:1203.1484 [hep-ph]].

\bibitem{Balantekin:2011ft} 
  A.~B.~Balantekin and A.~Malkus,
  Phys.\ Rev.\ D {\bf 85}, 013010 (2012)
  [arXiv:1109.5216 [hep-ph]].

\bibitem{Kostelecky:1993ys}
  V.~A.~Kostelecky and S.~Samuel,
  Phys.\ Rev.\  D {\bf 49}, 1740 (1994).

\bibitem{Abazajian:2002qx}
  K.~N.~Abazajian, J.~F.~Beacom and N.~F.~Bell,
  %
  Phys.\ Rev.\ D {\bf 66}, 013008 (2002)
  [arXiv:astro-ph/0203442].

\bibitem{Sawyer:2005jk} 
  R.~F.~Sawyer,
  Phys.\ Rev.\ D {\bf 72}, 045003 (2005)
  [hep-ph/0503013].

\bibitem{Duan:2008fd}
  H.~Duan, G.~M.~Fuller and Y.~Z.~Qian,
  J.\ Phys.\ G {\bf 36}, 105003 (2009)
  [arXiv:0808.2046 [astro-ph]].

\bibitem{Balantekin:2009dy} 
  A.~B.~Balantekin,
  \emph{Nucl.\ Phys.\ A} \textbf{844}, 14C-18C (2010)
  [arXiv:0910.1814 [nucl-th]].

\bibitem{Pehlivan:2010zz} 
  Y.~Pehlivan, T.~Kajino, A.~B.~Balantekin, T.~Yoshida and T.~Maruyama,
  \emph{AIP Conf.\ Proc.}  \textbf{1269}, 189-194 (2010).

\bibitem{Richardson1}
  R.W. Richardson,   Phys. Lett. \textbf{3} (1963) 277.
  
 \bibitem{Gaudin1}
  M. Gaudin, 
  ``Diagonalisation D'une Classe D'Hamiltoniens de Spin,'' 
  J. Physique \textbf{37}(1976), 1087.

\bibitem{Gaudin2}
  M. Gaudin, 
  ``La Fonction d'onde de Bethe,'' 
  Collection du Commissariat a l'\'{e}nergie atomique, Masson, Paris, 1983.

\bibitem{yuzb} 
 A.A. Yuzbashyan, B.L. Altshuler, V.B. Kuznetsov, and V.E. Enolskii, J. Phys. A: Math. Gen. {\bf 38}, 7831 (2005).   

\bibitem{Pehlivan:2011hp} 
  Y.~Pehlivan, A.~B.~Balantekin, T.~Kajino and T.~Yoshida,
  \emph{Phys.\ Rev.\ D} \textbf{84}, 065008 (2011)
  [arXiv:1105.1182 [astro-ph.CO]].

\bibitem{Raffelt:2011yb} 
  G.~G.~Raffelt,
  Phys.\ Rev.\ D {\bf 83}, 105022 (2011)
  [arXiv:1103.2891 [hep-ph]].

\bibitem{Duan:2007bt} 
  H.~Duan, G.~M.~Fuller, J.~Carlson and Y.~-Q.~Zhong,
  Phys.\ Rev.\ Lett.\  {\bf 99}, 241802 (2007)
  [arXiv:0707.0290 [astro-ph]].

\bibitem{Dasgupta:2009mg} 
  B.~Dasgupta, A.~Dighe, G.~G.~Raffelt and A.~Y.~.Smirnov,
  Phys.\ Rev.\ Lett.\  {\bf 103}, 051105 (2009)
  [arXiv:0904.3542 [hep-ph]].

\bibitem{Galais:2011gh} 
  S.~Galais and C.~Volpe,
  Phys.\ Rev.\ D {\bf 84}, 085005 (2011)
  [arXiv:1103.5302 [astro-ph.SR]].

\bibitem{Banerjee:2011fj} 
  A.~Banerjee, A.~Dighe and G.~Raffelt,
  Phys.\ Rev.\ D {\bf 84}, 053013 (2011)
  [arXiv:1107.2308 [hep-ph]].

\bibitem{Balantekin:2003ip} 
  A.~B.~Balantekin and G.~M.~Fuller,
  J.\ Phys.\ G G {\bf 29}, 2513 (2003)
  [astro-ph/0309519].

\bibitem{Cherry:2012zw} 
  J.~F.~Cherry, J.~Carlson, A.~Friedland, G.~M.~Fuller and A.~Vlasenko,
  arXiv:1203.1607 [hep-ph].

\end{thebibliography}
\end{document}